\documentclass[aps,pra,reprint,superscriptaddress,amsmath,nofootinbib]{revtex4-1}
\usepackage{bm,amsmath}
\usepackage{ifpdf}
\usepackage{siunitx}
\usepackage{braket}
\usepackage{comment}

\usepackage{graphicx}
\usepackage[colorlinks=true,urlcolor=blue,citecolor=blue,linkcolor=blue,breaklinks=true]{hyperref}
\input glyphtounicode
\usepackage{txfonts}
\usepackage{xcolor}

\graphicspath{{figs/}}

\DeclareGraphicsExtensions{.pdf}

\begin{document}

\title{Nonreciprocal Landau-Zener tunneling}

\author{Sota~Kitamura}
\affiliation{Department of Applied Physics, The University of Tokyo, Hongo, Tokyo, 113-8656, Japan}

\author{Naoto~Nagaosa}
\affiliation{Department of Applied Physics, The University of Tokyo, Hongo, Tokyo, 113-8656, Japan}
\affiliation{RIKEN Center for Emergent Matter Sciences (CEMS), Wako, Saitama, 351-0198, Japan}

\author{Takahiro~Morimoto}
\affiliation{Department of Applied Physics, The University of Tokyo, Hongo, Tokyo, 113-8656, Japan}

\begin{abstract}
Application of strong dc electric field to an insulator leads to quantum tunneling of electrons from the valence band to the conduction band, which is a famous nonlinear response known as Landau-Zener tunneling. One of the growing interests in recent studies of nonlinear responses is nonreciprocal phenomena where transport toward the left and the right differs. 
Here, we theoretically study Landau-Zener tunneling in noncentrosymmetric systems, i.e., the crystals without spatial inversion symmetry.
A generalized Landau-Zener formula is derived taking into account the geometric nature of the wavefunctions. 
The obtained formula shows that nonreciprocal tunneling probability originates from the difference in the Berry connections of the Bloch wavefunctions across the band gap, i.e., shift vector. 
We also discuss application of our formula to tunneling in a one-dimensional model of a ferroelectrics. 
\end{abstract}

\date{\today}

\maketitle

\section{Introduction}
\begin{figure}[t]
	\begin{centering}
		\includegraphics[width=\linewidth]{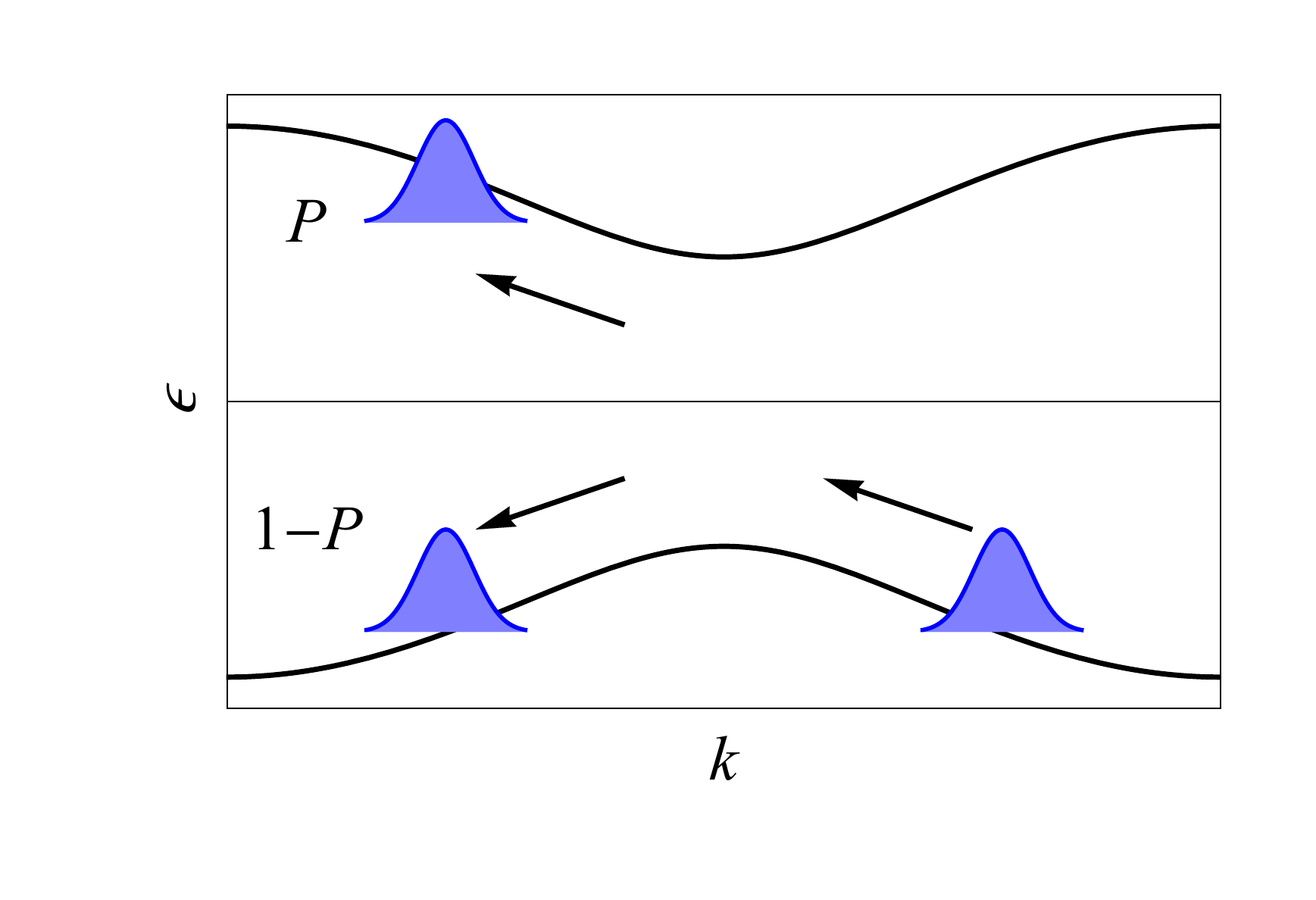}
		\par\end{centering}
	\caption{Schematics of Landau-Zener tunneling. A wave packet driven by an electric field can tunnel into the conduction band with transition probability $P$.
	}
	\label{fig: band}
\end{figure}

Tunneling phenomenon is one of the most remarkable and unique consequences 
of the wave nature of particles in quantum mechanics, 
where a particle can penetrate through classically forbidden regions. 
In solids, the quantum mechanical wavefunctions of electrons form 
the band structure separated by the energy gaps, 
and the tunneling can occur between these bands when an electric field is applied. 
This is called Zener tunneling through the energy gap and has been actively studied~\cite{LandauLifshitz,Zener32,Davis1976,Dykhne1962,George1974,Berry1990,Joye1991-1,Joye1991-2,Wu00,Liu02,Saito07,Kayanuma08,Oka2012}. 
A concise formula, i.e., Landau-Zener formula~\cite{LandauLifshitz,Zener32}, has been obtained for a model Hamiltonian 
describing the two-band system as  
\begin{align}
H = \begin{pmatrix}v\hbar k&\delta\\\delta&-v\hbar k
\end{pmatrix},
\end{align}
where $\pm v\hbar k $ are the energy dispersions and 
$2 \delta$ is the energy gap. 
Under an external electric field $E$, 
the wavenumber $k$ is accelerated as $\hbar{\dot k} = -eE$ as shown in Fig.~\ref{fig: band}. 
The transition probability from the lower band to the upper band reads
\begin{align}
P=\exp\left(-\frac{\pi\delta^2}{e\hbar Ev}\right),
\label{eq:conventional-LZ}
\end{align}
which is essentially singular with respect to $E$ showing the nonperturbative nature of the quantum tunneling. 

At a pn-junction of semiconductors, the tunneling shows an asymmetric behavior, which is utilized 
as a tunneling diode for rectifying devices~\cite{Esaki58}. 
Because of the broken inversion symmetry, 
the tunneling probability, and hence, the I-V characteristics depend strongly 
on the direction of the electric field $E$. 
For the uniform bulk crystal, however, 
the asymmetry in the Zener tunneling probability is a highly nontrivial issue 
even when the crystal lacks the inversion symmetry. 
This can be seen in the band dispersion $\varepsilon_n (k)$ ($n$: band index); 
the relation $\varepsilon_n (k) = \varepsilon_n (-k)$ holds due to the time-reversal symmetry 
even in the absence of the spatial inversion symmetry. 
Therefore, the inversion symmetry is rather hidden in wave mechanics~\cite{Tokura-Nagaosa18}.
Intuitively, the extended wave state is rather insensitive to 
the broken inversion symmetry compared with the localized wave-packet. 
Therefore, a fundamental question is how the nonreciprocal behavior, i.e., 
the asymmetry between the opposite direction of the electric field $E$, 
is realized in the tunneling processes of the bulk crystals, reflecting the wave nature of the electrons.   
This is also an important issue in terms of device applications;
Ferroelectric random access memory (FeRAM) utilizes nonreciprocal current response at the time of read-out operations of recorded polarization direction \cite{Han13}, while its working mechanism has not been fully understood so far.

\begin{figure*}[t]
	\begin{centering}
		\includegraphics[width=.8\linewidth]{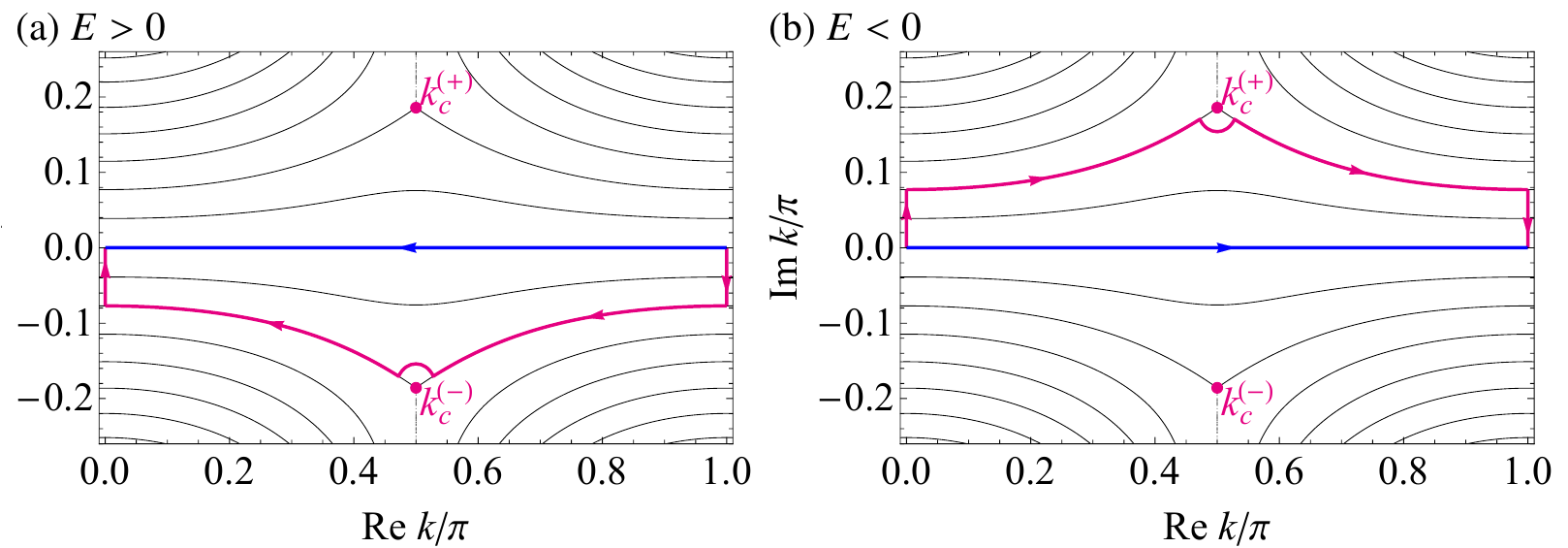}
		\par\end{centering}
	\caption{Complex $k$ plane that governs the tunneling process.
		The branching point $k_c$ is a point where the band gap vanishes in the complex plane.
		Cauchy's theorem allows to deform the original integration path on the real axis (blue) into the contour in the complex plane that passes through the branching point (magenta). 
		The direction of the original path is different for (a) $E>0$ and (b) $E<0$, and the choice of the branching point is also different.
	}
	\label{fig: k plane}
\end{figure*}

The nonreciprocal phenomena in noncentrosymmetric crystals have been extensively studied in these days,
including both the dc transport~\cite{Rikken01,RikkenCNT,RikkenNature,RikkenSi,Pop14,Wakatsuki17} and photo-excited current~\cite{Grinberg13,Nie,Shi,deQuilettes,Weber08,Kraut,Belinicher,Sipe,Young-Rappe,Cook17,Morimoto-Nagaosa16}. 
In particular, the NO-GO theorem has been proposed for the nonreciprocal transport of independent particles induced by the static electric field, in terms of a perturbative expansion with respect to $E$~\cite{Morimoto18}. 
Instead, the interacting electrons can show nonreciprocal dc transport in a perturbative treatment. 
On the other hand, this theorem does not apply for the photocurrent induced by 
the light irradiation which induces the inter-band transitions, 
which is called shift current. 
The shift current is formulated in terms of the Berry connection of the Bloch wavefunctions, 
which correspond to the intracell coordinates of the electrons~\cite{Sipe,Young-Rappe,Cook17,Morimoto-Nagaosa16,Nagaosa-Morimoto17}. 
The optical transition causes the shift in the intracell coordinates, 
i.e., shift vector, since intracell coordinates are generally different for the valence and conduction bands in noncentrosymmetric crystals.
The steady pumping of polarization of photoexcited electron-hole pairs results in the dc photocurrent. 
Therefore, it is concluded that the wavefunctions encode the information of the noncentrosymmetry 
in sharp contrast to the energy dispersion. 
In fact, the Berry phase becomes zero (or trivial) when the system preserves both the inversion and time-reversal symmetries. 

As discussed above, the tunneling is a nonperturbative effect, 
and cannot be captured by the perturbative expansion with respect to $E$. 
Hence, it is possible that the nonreciprocal nature appears in the Landau-Zener tunneling 
even in the independent particle approximation.  
Indeed, we show below that this is the case by deriving the generalized Landau-Zener formula 
including the shift vector, i.e., the information of the Bloch wavefunctions.

\section{Tunneling formula with a shift vector}
Let us consider a time evolution of a system under a slow change of
parameters. In particular, here we focus on a change of momentum $k$
under a DC electric field, $k\rightarrow k(t)=k-eEt/\hbar$. It is well known
that the solution of the time-dependent Schr\"odinger equation in
the adiabatic limit is given by snapshot eigenstates
\begin{align}
H(t)|n,k(t)\rangle=\varepsilon_{n}(t)|n,k(t)\rangle
\end{align}
multiplied by dynamical and Berry phase factors [see Eq.~(\ref{eq:psi})]. The diabatic correction
is derived from the transition dipole matrix elements. To see
this, let us expand a state vector $|\Psi\rangle$ by the adiabatic
solutions as 
\begin{align}
|\Psi\rangle=\sum_{n}a_{n}(t)e^{-i\int_{t_{0}}^{t}dt_{1}[\varepsilon_{n}(t_{1})+eEA_{nn}(t_{1})]/\hbar}|n,k(t)\rangle,
\label{eq:psi}
\end{align}
where $A_{nm}(t)=i\langle n,k(t)|\partial_{k}|m,k(t)\rangle$ is the Berry connection. (We note that the ``off-diagonal'' Berry connections for $n \neq m$ correspond to transition dipole matrix elements.)
With paying attention in dealing with the Berry phase factor,
we can reduce the time-dependent Schr\"odinger equation $i\hbar\partial_{t}|\Psi\rangle=H(t)|\Psi\rangle$
to 
\begin{align}
i\partial_{t}a_{n}(t)=\frac{eE}{\hbar}\sum_{m\neq n}|A_{nm}(t)|e^{i\int_{t_{0}}^{t}dt_{1}[\varepsilon_{n}-\varepsilon_{m}+eER_{nm}]/\hbar+i\arg A_{nm}(t_{0})}a_{m}(t),
\label{eq:amp-eom}
\end{align}
with 
\begin{align}
R_{nm}=A_{nn}-A_{mm}-\partial_{k}\arg A_{nm}.
\end{align}
See Appendix~\ref{sec:adiabatic} for details.
Here we have used $\hbar\partial_{t}=-eE\partial_{k}$ for $|n,k(t)\rangle$
and $\arg A_{nm}(t)$. $R_{nm}$ is so called shift vector 
which is a gauge-invariant object describing the polarization difference
between two bands $n,m$~\cite{Sipe,Young-Rappe,Cook17,Morimoto-Nagaosa16,Nagaosa-Morimoto17,Sinitsyn}. 
Specifically, the shift vector can be interpreted as the spatial shift of the wave packets between the valence and conduction bands, since the Berry connections $A_{nn}$ correspond to the intracell coordinates of the Bloch electrons. 
It is known that the shift vector plays an important role in bulk photogalvanic effect in noncentrosymmetric crystals, so called shift current.
The fact that $R_{nm}$ appears in Eq.~(\ref{eq:amp-eom}) is usually overlooked since $A_{nn}=0$ is assumed in many cases. 
A similar geometric contribution has also been discussed as the quantum geometric potential~\cite{Wu2008,Wu2018}, in the context of the adiabatic condition.

Let us focus on a tunneling process between two bands, $n=\pm$ with
$a_{-}(t_{0})=1$, $a_{+}(t_{0})=0$. Our goal is to derive the tunneling probability $P=|a_+(t)|^2$ after one cycle of the Bloch oscillation.
For simplicity, we consider
only the first-order correction w.r.t. $|A_{+-}|$ here. By integrating
Eq.~(\ref{eq:amp-eom}) and using it recursively, we obtain 
\begin{align}
a_{+}(t)=&ie^{i\arg A_{+-}(t_{0})}\notag\\ &\times\int_{k_0}^{k(t)}dk_{1}\left|A_{+-}\right|\exp\left[-i\int_{k_{0}}^{k_{1}}dk_{2}\left(\frac{\varepsilon_{+}-\varepsilon_{-}}{eE}+R_{+-}\right)\right]
\label{eq:amp-1st}
\end{align}
with $k_0=k(t_0)$, as we detail in Appendix~\ref{sec:adiabatic}.

A two-band Hamiltonian can be represented as $H=\bm{d}(k)\cdot\bm{\sigma}$
with $\bm{\sigma}$ being Pauli matrices (when we subtract a constant
energy shift). The quantities necessary for the evaluation of the tunneling
amplitude are given as 
\begin{gather}
\varepsilon_{+}-\varepsilon_{-}=2\sqrt{\bm{d}^2},\\
|A_{+-}|=\frac{\sqrt{(\bm{d}\times\partial_{k}\bm{d})^2}}{2\bm{d}^2},\label{eq:asym-bc}\\
R_{+-}=\frac{(\bm{d}\times\partial_{k}\bm{d})\cdot(\partial_{k}^{2}\bm{d})}{(\bm{d}\times\partial_{k}\bm{d})^{2}}\sqrt{\bm{d}^2}.
\label{eq:shift-vector}
\end{gather}

In order to evaluate the integral in an asymptotic manner, we employ
the Dykhne-Davis-Pechukas (DDP) method~\cite{Dykhne1962,Davis1976} in accordance with Ref.~\onlinecite{Davis1976}.
Namely, we perform the integral by means of contour integration in
the complex plane. The contour of the integral, which is originally
the real axis [blue lines in Fig.~\ref{fig: k plane}], 
can be deformed within an analytic region, thanks to the Cauchy's integral theorem~\cite{Dykhne1962,Pokrovskii}. 

This treatment is advantageous since one can utilize a (complex) branching point $k_c$ 
where the energy gap vanishes $\bm{d}(k_{c})^2=0$
[Such a point is indeed a branching point when the Hamiltonian is analytic,
as $\varepsilon_{+}-\varepsilon_{-}\propto(k-k_c)^{1/2}$ in the vicinity of $k_c$].
This point essentially governs the tunneling process between the two bands:
Since the prefactor $|A_{+-}|$ diverges as we approach $k_c$ [see Eq. (\ref{eq:asym-bc})], 
only this divergent part contributes to the asymptotic value of the integral, 
when the integration path is deformed to pass through the vicinity of the branching point $k_c$.

We show the integration path by magenta lines in Fig.~\ref{fig: k plane}.
The main part of the contour is one along which the absolute value of the exponential factor is constant (i.e., the imaginary
part of the $k_{2}$ integral in Eq.~(\ref{eq:amp-1st}) is constant).
This contour passes through the branching point $k_c$, but we make a detour around it 
since $k_c$ itself is a singular point of the integrand.
Due to the divergence mentioned above, this detoured part contributes dominantly against the main part.
While the branching points appear in a pairwise manner $(k_c,k_c^\ast)$, 
we need to choose one of them such that the exponential factor becomes smaller than unity in accordance with the detailed derivation given in Appendix~\ref{sec:ddp}.
We note that the integral on the first and last vertical lines have finite but small contributions in general. They share the same absolute value (that is perturbative in $E$) but their phases are different. This leads to a small oscillation in the tunneling amplitude with respect to $E$, on top of the nonperturbative contribution from the branching point $k_c$. Since we are interested in the nonperturbative contribution, we neglect those perturbative corrections in the following.

In a generic situation, one can assume that the leading order term of 
$\bm{d}^{2}$, $(\partial_{k}\bm{d})^{2}$
and $(\bm{d}\times\partial_{k}\bm{d})\cdot(\partial_{k}^{2}\bm{d})$
in the expansion around $k_{c}$ is given as $\bm{d}^{2}\sim i\alpha(k-k_{c})$,
$(\partial_{k}\bm{d})^{2}\sim\beta$, and $(\bm{d}\times\partial_{k}\bm{d})\cdot(\partial_{k}^{2}\bm{d})\sim\eta$,
respectively. By evaluating the detoured part of the integral [circular arc around $k_c$ in Fig.~\ref{fig: k plane}] with
these expanded forms, we arrive at 
\begin{align}
P\sim\exp\left[2\text{Im}\int_{k_{0}}^{k_{c}}dk_{2}\left(\frac{\varepsilon_{+}-\varepsilon_{-}}{eE}+R_{+-}\right)\right],
\label{eq:new-formula}
\end{align}
as we describe in Appendix~\ref{sec:ddp}. We note that there actually is a prefactor $(\pi/3)^2 \sim 1.1$,
which appears because we have evaluated the probability within the first order of the adiabatic perturbation.
However, we dropped this because it is known~\cite{Davis1976} that the prefactor becomes exactly 1 
if we take into account all order contributions in the adiabatic perturbation (for details, see Appendix~\ref{sec:ddp}).
We also note that the integral in Eq.~(\ref{eq:new-formula}) is invariant against replacing the lower limit $k_0$ to an arbitrary $k$ on the real axis (usually it is set to the gap minimum).

The obtained formula, Eq.~(\ref{eq:new-formula}), includes the geometric correction
described by the shift vector. This contribution is absent in the original DDP formula, because they
assumed that the $2\times2$ Hamiltonian is real; Under this assumption
$\bm{d}$ is a two-dimensional vector, where $(\bm{d}\times\partial_{k}\bm{d})\cdot(\partial_{k}^{2}\bm{d})=0$
always holds. 
On the other hand, Refs.~\onlinecite{Berry1990,Joye1991-1,Joye1991-2} dealt with
a generic $2\times2$ Hamiltonian, so that the geometric correction
indeed appeared in their study. However, their calculation was done
in a particular gauge, and the obtained expression is not gauge-invariant.
Comparison of the formula is given in Appendix~\ref{sec:comparison}.
Our result is expressed in terms of gauge-invariant quantities, the shift vector, so
that it provides much clearer understanding on the physical interpretation
of the correction and remarkable physical consequences due to it.

The shift vector $R_{nm}$ plays a crucial role in the nonlinear transport of inversion-broken systems.
It satisfies $R_{nm}(k)=R_{nm}(-k)$ when the system is time-reversal symmetric, while $R_{nm}(k)=-R_{nm}(-k)$ when the system is inversion symmetric.
Thus when the system has both symmetries, there is no correction to the tunneling probability; 
This is consistent with the fact that one can make the Hamiltonian real in such cases.
A nontrivial result thus can appear when either symmetry is broken.
In particular, a qualitatively new phenomenon appears when the inversion symmetry is broken:
When we change the sign of the electric field $E$, we need to change the choice of the branching point 
from $k_c$ to $-k_c$ in order to have a correct result $P\le1$ with the formula (\ref{eq:new-formula})
[See Figs.~\ref{fig: k plane}(a) and (b)].
Under this alternation, the shift vector contribution is invariant in the time-reversal broken system, 
which leads to a simple correction of the probability independent of the field strength/direction.
On the other hand, when the inversion symmetry is broken, the exponent of the shift vector correction is odd under this alternation,
so that it leads to an exponentially-large difference in the tunneling probability when the direction of the electric field is reverted.

\begin{figure*}
\begin{centering}
\includegraphics[width=\linewidth]{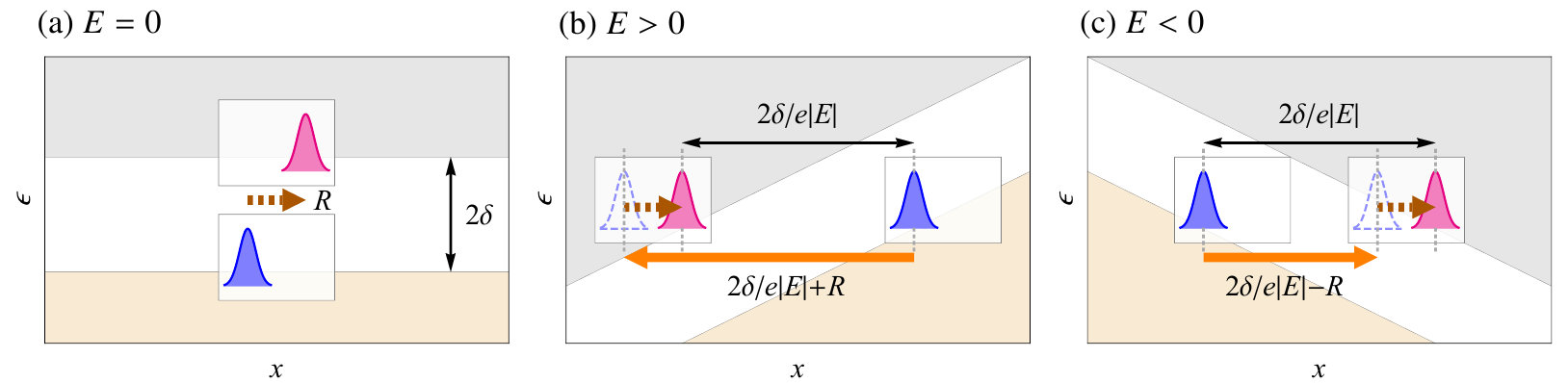}
\par\end{centering}
\caption{Zener tunneling in inversion broken systems. 
(a) Real space picture in the absence of the electric field $E$. The positions of the wave packets in the two bands are shifted in the unit cell by the shift vector $R$. 
(b,c) Real space pictures of two bands in the presence of $E$.
The shift in the intracell coordinate results in the different tunneling depth depending on the direction of $E$ by the shift vector $R$.
Namely, the shift of the wave packet by $R$ upon the interband transition effectively modifies the tunneling depth as $2\delta/eE+R$.
}
\label{fig: tunneling with shift vector}
\end{figure*}

The physical meaning of the shift vector as an intracell coordinate provides an intuitive understanding of its role in the tunneling process, as follows.  
The tunneling process can be interpreted as a propagation of a wave packet thorough a classically forbidden region in real space, 
which has a thickness of $2\delta/eE$ as drawn in Fig.~\ref{fig: tunneling with shift vector}.
Now the noncentrosymmetric system has an internal degree of freedom, the intracell coordinate, whose difference between two bands is 
represented by the shift vector [See Fig.~\ref{fig: tunneling with shift vector}(a)].
Thus the tunneling from the lower band to the upper band induces a spontaneous shift of the wave packet by $R$, which effectively modifies the tunneling depth as $2\delta/eE+R$.
Since the direction of the shift vector is intrinsically determined by the underlying crystal, 
this modification of the tunneling depth by $R$ contributes in a constructive/destructive manner in accordance with the direction of the bias,
as shown in Figs.~\ref{fig: tunneling with shift vector}(b) and (c). 
As can be seen in the tunneling formula (\ref{eq:new-formula}), this modification of the tunneling depth is essentially governed by the shift vector $R_{+-}$ (defined with the Berry connection difference between the two bands) at the band gap minimum.

\section{Application to Rice-Mele model}
Let us see the emergence of the nonreciprocal (direction-dependent) tunneling probability using the Rice-Mele model (with the lattice constant $a=1$)
\begin{align}
H=\begin{pmatrix}m & t\cos k-i\delta t\sin k\\
t\cos k+i\delta t\sin k & -m
\end{pmatrix},
\label{eq:RM}
\end{align}
a prototypical example of an inversion-broken (polar) system. Indeed, this
model has a finite shift vector 
\begin{align}
R_{+-}=\frac{mt\delta t\sqrt{t^{2}\cos^{2}k+\delta t^{2}\sin^{2}k+m^{2}}}{m^{2}(\delta t^{2}\cos^{2}k+t^{2}\sin^{2}k)+t^{2}\delta t^{2}}.
\end{align}
Since the integrand of Eq.~(\ref{eq:new-formula}) for the present
system is invariant under $k\rightarrow k\pm\pi$, let us consider a
slow parameter change $k:\:0\rightarrow-\pi\text{sgn}(E)$. The branching point
of the present model is given as  
\begin{align}
k_{c}^{(\pm)}=\frac{2n+1}{2}\pi\pm i\tanh^{-1}\sqrt{\frac{\delta t^{2}+m^{2}}{t^{2}+m^{2}}}
\end{align}
with $n\in\mathbb{Z}$.
By evaluating the generalized formula (\ref{eq:new-formula}) with $k_{c}=k_{c}^{(-\text{sgn}(E))}$, 
we obtain \begin{widetext} 
\begin{align}
P & =\exp\left[-\frac{4\sqrt{t^{2}+m^{2}}}{e|E|}\left(K(\gamma)-E(\gamma)\right)-\frac{2m\text{sgn}(E)}{t\delta t\sqrt{t^{2}+m^{2}}}\left(t^{2}K(\gamma)-(t^{2}-\delta t^{2})\Pi(\frac{\delta t^{2}}{t^{2}},\gamma)\right)\right]\label{eq:RM-P0}\\
 & \sim\exp\left[-\frac{\pi(\delta t^{2}+m^{2})}{e|E|t}-\text{sgn}(E)\frac{\pi m\delta t}{2t^2}\right]\quad(\delta t,m\ll t),
 \label{eq:RM-P}
\end{align}
\end{widetext} where $K(\gamma)$, $E(\gamma)$, and $\Pi(n,\gamma)$
are the complete elliptic integrals with $\gamma=\sqrt{(\delta t^{2}+m^{2})/(t^{2}+m^{2})}$
being the elliptic modulus. 
In the last line, we recover the conventional exponent Eq.~(\ref{eq:conventional-LZ}) for the $|E|^{-1}$ term ($\delta\rightarrow\sqrt{\delta t^2+m^2}$ and $\hbar v\rightarrow t$). The shift vector correction is represented as $-\text{sgn}(E)\times (\pi/2)|\text{Im}k_c|\times R_{+-}$ evaluated at $k=\pi/2$. 
Remarkably, as we have mentioned, the non-zero shift vector not just provides an exponentially-large correction, 
but also a strong nonreciprocity via $\text{sgn}(E)$.

\begin{figure}
\begin{centering}
\includegraphics[width=\linewidth]{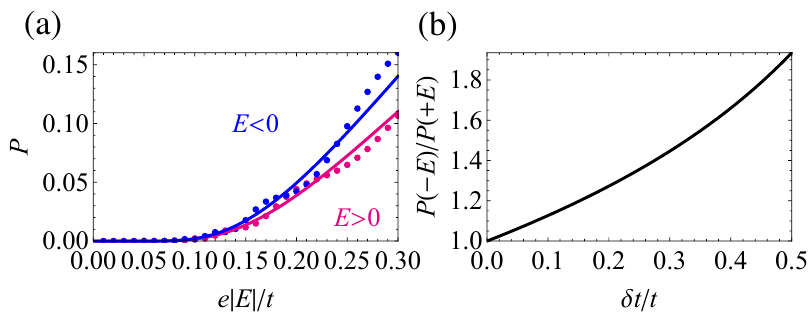}
\par\end{centering}
\caption{(a) Nonreciprocal tunneling probability $P(E)$ in Rice-Mele model. Solid lines represent $P(E)$ from the generalized Landau-Zener formula, and dots represent $P(E)$ obtained from numerical simulations of time-dependent Schr\"odinger equation.
We used the parameters $(\delta t,m)=(0.2t,0.4t)$. 
(b) Nonreciprocity ratio $P(-E)/P(+E)$ plotted as a function of alternating hopping amplitude $\delta t$ that controls the magnitude of inversion breaking.
We set $m=0.4t$.
}
\label{fig: I-V for RM model}
\end{figure}

Using the obtained generalized Landau-Zener formula, we show the tunneling probability $P$ in Rice-Mele model in Fig.~\ref{fig: I-V for RM model}.
Plots of $P(E)$ as a function of the applied electric field $E$ (Fig.~\ref{fig: I-V for RM model}(a)) show a well-known nonpertrubative behavior at $E=0$ and a good agreement with tunneling probabilities obtained by numerically solving time-dependent Schr\"odinger equation (shown in dots).
We note that the oscillation with a small amplitude seen in the numerical results arises from the perturbative corrections due to the integral along the first and last vertical lines of Fig.~\ref{fig: k plane}.
In addition, the plots show that tunneling probabilitites differ depending on the direction of $E$ (the sign of $E$). 
This direction dependence arises from the nonzero shift vetor and shows almost proportionality to the value of $R_{+-}$ at the band gap minimum (on the real axis of $k$).
Figure~\ref{fig: I-V for RM model}(b) shows the nonreciprocity ratio 
$P(-E)/P(+E)$
(the ratio of tunneling probabilities for negative and positive $E$),
which quantifies the strength of nonreciprocity as a rectifying device,
as a function of the strength of alternating hopping $\delta t$ that introduces inversion symmetry breaking.
We find that the nonreciprocity ratio $P(-E)/P(+E)$ grows monotonically, as we increase $\delta t$ and the effect of inversion symmetry breaking gets stronger.
For small $\delta t$, the nonreciprocity ratio is linearly proportinoal to $\delta t$. 
In particular, large nonreciprocity of $P(-E)/P(+E) \sim 2$ can be achieved for a feasible value of $\delta t \sim 0.5t$, which indicates that Landau Zener tunneling in noncentrosymmetric crystals is able to realize strong nonreciprocal functinoality.

\section{Nonreciprocal tunneling in a continuous model}
\begin{figure}[t]
\begin{centering}
\includegraphics[width=\linewidth]{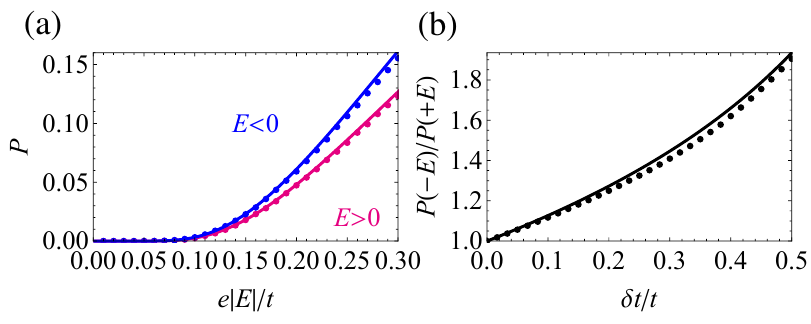}
\par\end{centering}
\caption{\label{fig:continuous}
Nonreciprocal tunneling in the continuous model. (a) Nonreciprocal tunneling probability $P(E)$. 
Solid lines represent $P(E)$ from the generalized Landau-Zener formula, and dots represent $P(E)$ obtained from numerical simulations of time-dependent Schr\"odinger equation.
We used the parameters $(\delta t,m)=(0.2t,0.4t)$. 
(b) Nonreciprocity ratio $P(-E)/P(+E)$ plotted as a function of alternating hopping amplitude $\delta t$ that controls the magnitude of inversion breaking.
We set $m=0.4t$.
}
\end{figure}

As we have mentioned, the small-amplitude oscillation of the tunneling probability seen
in Fig.~\ref{fig: I-V for RM model}(a) is derived from the first and last vertical segments
of the deformed contour shown in Fig.~\ref{fig: k plane}. This contribution survives
due to the finite bandwidth, and should vanish when the energy gap
is infinite at the initial and final states. To verify this, we consider
a Hamiltonian in a continuous space,
\begin{equation}
H=\begin{pmatrix}m\sqrt{1+k^{2}} & -tk-i\delta t\\
-tk+i\delta t & -m\sqrt{1+k^{2}}
\end{pmatrix},\label{eq:continuous}
\end{equation}
whose low-energy structure coincides with that of the Rice-Mele model~(\ref{eq:RM}), while the energy spectrum is unbounded. 
The generalized formula (\ref{eq:new-formula}) for this Hamiltonian reads 
\begin{align}
P=\exp &\Bigg[{-\pi(\delta t^{2}+m^{2})\over e|E|\sqrt{t^{2}+m^{2}}}
\nonumber \\
& 
-\frac{2m\text{sgn}(E)}{t\delta t\sqrt{t^{2}+m^{2}}}\left(t^{2}K(\gamma)-(t^{2}-\delta t^{2})\Pi(\frac{\delta t^{2}}{t^{2}},\gamma)\right)\Bigg].
\end{align}
It is worth noting that the nonreciprocal part ($\propto\text{sgn}(E)$)
is identical to that in Eq.~(\ref{eq:RM-P0}),
 while the reciprocal part is modified from Eq.~(\ref{eq:RM-P0}). 
We compare the formula with numerically
computed tunneling amplitudes and nonreciprocity ratios in Fig.~\ref{fig:continuous}.
The numerical result no longer shows oscillation with respect to $E$ from the perturbative correction and shows better agreement with the generalized Landau-Zener formula.

\section{Discussions}
Here we further consider the role of time-reversal symmetry $T$ in the nonreciprocal responses. 
In the context of magnetochiral anisotropy~\cite{Rikken01,RikkenCNT,RikkenNature,RikkenSi,Pop14}, it has been discussed that the nonreciprocal transport requires the broken $T$ in addition to the broken inversion symmetry. 
This can be understood intuitively that the reversal of time corresponds to that of the current direction when there is no dissipation. 
The NO-GO theorem in Ref.~\onlinecite{Morimoto18} indeed shows that the current proportional to the square of the electric field is forbidden in non-interaction systems with dc electric field, when the $T$-symmetry is preserved. 
However, it has been revealed that this is not the case when the inter-band transitions and the associated absorption of energy occur due to ac $E$-field, where the shift current is induced. 
The Zener tunneling studied in the present paper can be regarded as the ``inter-band" transition under dc $E$-field 
due to the tunneling, and hence the NO-GO theorem, 
which is based on the adiabatic assumption that there occurs no inter-band transition, does not apply. 
This is related to the non-analytic and non-perturbative nature of the tunneling probability, which cannot be expanded in $E$. 
In a more general situation in the plane of the frequency $\omega$ and the strength $E$ of the electric field, 
there are several different regions as indicated by Fig.~17 of Ref.~\onlinecite{Aoki2014}. 
In this respect, the shift current and Zener tunneling are two limiting cases, 
i.e., ac limit of weak $E$-field and dc limit of strong $E$-field, 
and the crossover between these two corresponds to the Keldysh line and 
how the nonreciprocal responses behaves in this plane is an interesting problem to be studied in the future. 
Note that from the viewpoint of the symmetry, the requirement for the nonreciprocal Zener tunneling is 
the same as that for the shift current. 
Another unique feature of the nonreciprocal Zener tunneling is that the ratio of the tunneling currents 
for the two directions is independent of the electric field $E$ and is of the order of unity as indicated in Eq.~(\ref{eq:RM-P}). 

It is an interesting future problem to investigate further geometric corrections. 
There might be new effects that do not emerge until the higher-order perturbation is taken into account~\cite{Nagaosa2010,Xiao2019}, 
which can lead to even larger nonreciprocity. 
Since the geometric correction discussed in the present study appears in a combination of $ER_{mn}$, 
and we have evaluated the nonperturbative contribution within the leading order of $E$ (for the prefactor), 
such new effects might be found in the presence of strong $E$ field.

In the present paper, we considered the spinless model. 
Incorporating the spin degrees of freedom and spin-orbit interaction, 
one can expect the nonreciprocal charge and spin tunneling currents. 
These phenomena can offer novel mechanisms for spin diode or switchable diode by a magnetic field. 
The large spin polarization of the electrons transmitted through 
the DNA molecules that has been observed experimentally~\cite{Gohler11} might be understood from this point of view~\cite{Matityahu16}.

\begin{acknowledgments}

We thank J. Wu and T. Oka for discussions at an early stage of this work which gave us an important implication on the role of the geometrical phase in tunneling phenomena.
We also thank Yoshinori Tokura for fruitful discussions.
This work was supported by JST CREST (JPMJCR19T3) (SK, TM), The University of Tokyo Excellent Young Researcher Program and JST PRESTO (JPMJPR19L9) (TM),  JST CREST (JPMJCR1874 and JPMJCR16F1) and JSPS KAKENHI (18H03676 and 26103006)(NN).

\end{acknowledgments}

\setcounter{section}{0}
\renewcommand{\thesection}{}%

\begin{widetext}
\section*{Appendix}
\subsection{Adiabatic perturbation theory\label{sec:adiabatic}}
Equations~(\ref{eq:amp-eom}, \ref{eq:amp-1st}) in the main text are derived 
using the adiabatic perturbation theory as follows.
By inserting Eq.~(\ref{eq:psi}) to the time-dependent Schr\"odinger equation,
we obtain 

\begin{equation}
\sum_{m}e^{-i\int_{t_{0}}^{t}dt_{1}[\varepsilon_{m}(t_{1})+eEA_{mm}(t_{1})]/\hbar}\left\{ [i\hbar\partial_{t}a_{m}(t)+eEa_{m}(t)A_{mm}(t)]|m,k(t)\rangle-eEa_{m}(t)i\partial_{k}|m,k(t)\rangle\right\} =0.
\end{equation}
Here we have used $\hbar\partial_{t}|m,k(t)\rangle=-eE\partial_{k}|m,k(t)\rangle.$
By taking an inner product with $\langle n,k(t)|$, $\langle n,k(t)|m,k(t)\rangle=\delta_{nm}$
and $\langle n,k(t)|i\partial_{k}|m,k(t)\rangle=A_{nm}(t)$ leads
to
\begin{align}
i\partial_{t}a_{n}(t) & =\frac{eE}{\hbar}\sum_{m\neq n}A_{nm}(t)e^{i\int_{t_{0}}^{t}dt_{1}[\varepsilon_{n}(t_{1})+eEA_{nn}(t_{1})]/\hbar}e^{-i\int_{t_{0}}^{t}dt_{1}[\varepsilon_{m}(t_{1})+eEA_{mm}(t_{1})]/\hbar}a_{m}(t)\\
 & =\frac{eE}{\hbar}\sum_{m\neq n}|A_{nm}(t)|e^{i\int_{t_{0}}^{t}dt_{1}[\varepsilon_{n}-\varepsilon_{m}+eE(A_{nn}-A_{mm})]/\hbar+i\arg A_{nm}(t)}a_{m}(t)\\
 & =\frac{eE}{\hbar}\sum_{m\neq n}|A_{nm}(t)|e^{i\int_{t_{0}}^{t}dt_{1}[\varepsilon_{n}-\varepsilon_{m}+eE(A_{nn}-A_{mm})+\hbar\partial_{t}\arg A_{nm}]/\hbar+i\arg A_{nm}(t_{0})}a_{m}(t)\\
 & =\frac{eE}{\hbar}\sum_{m\neq n}|A_{nm}(t)|e^{i\int_{t_{0}}^{t}dt_{1}[\varepsilon_{n}-\varepsilon_{m}+eE(A_{nn}-A_{mm}-\partial_{k}\arg A_{nm})]/\hbar+i\arg A_{nm}(t_{0})}a_{m}(t).
\end{align}
This coincides with Eq.~(\ref{eq:amp-eom}).

By integrating Eq.~(\ref{eq:amp-eom}) from $t_{0}$ to $t$, we obtain a perturbative expansion in terms of the adiabatic perturbation theory, 
\begin{align}
a_{n}(t)  = & a_{n}(t_{0})-i\frac{eE}{\hbar}\int_{t_{0}}^{t}dt_{1}\sum_{m\neq n}|A_{nm}(t_{1})|e^{i\int_{t_{0}}^{t_{1}}dt_{2}[\varepsilon_{n}-\varepsilon_{m}+eER_{nm}]/\hbar+i\arg A_{nm}(t_{0})}a_{m}(t_{1}).\\
  = & a_{n}(t_{0})-i\frac{eE}{\hbar}\int_{t_{0}}^{t}dt_{1}\sum_{m\neq n}|A_{nm}(t_{1})|e^{i\int_{t_{0}}^{t_{1}}dt_{2}[\varepsilon_{n}-\varepsilon_{m}+eER_{nm}]/\hbar+i\arg A_{nm}(t_{0})}a_{m}(t_{0})\nonumber \\
   & +\left(-i\frac{eE}{\hbar}\right)^{2}\int_{t_{0}}^{t}dt_{1}\int_{t_{0}}^{t_{1}}dt_{2}\sum_{m\neq n}\sum_{l\neq m}|A_{nm}(t_{1})||A_{ml}(t_{2})| e^{i\int_{t_{0}}^{t_{1}}dt_{2}[\varepsilon_{n}-\varepsilon_{m}+eER_{nm}]/\hbar+i\int_{t_{0}}^{t_{2}}dt_{3}[\varepsilon_{m}-\varepsilon_{l}+eER_{ml}]/\hbar+i\arg A_{nm}(t_{0})+i\arg A_{ml}(t_{0})}a_{l}(t_{2})].
\end{align}
\end{widetext}
We then neglect the third term $O(|A|^{2})$ in the last line,
and set $n=+$, $m=-$. We obtain Eq.~(\ref{eq:amp-1st}) as 
\pagebreak
\begin{align}
a_{+}(t) & =-i\frac{eE}{\hbar}\int_{t_{0}}^{t}dt_{1}|A_{+-}(t_{1})|e^{i\int_{t_{0}}^{t_{1}}dt_{2}[\varepsilon_{+}-\varepsilon_{-}+eER_{+-}]/\hbar+i\arg A_{+-}(t_{0})}\\
 & =i\int_{k(t_{0})}^{k(t)}dk_{1}|A_{+-}(k_{1})|e^{-i\int_{k(t_{0})}^{k(t_{1})}dk_{2}[\frac{\varepsilon_{+}-\varepsilon_{-}}{eE}+R_{+-}]+i\arg A_{+-}(t_{0})},
\end{align}
since $a_{-}(t_{0})=1$ and $a_{+}(t_{0})=0$.

\subsection{Dykhne-Davis-Pechukas (DDP) method\label{sec:ddp}}
We evaluate Eq.~(\ref{eq:amp-1st}) as Eq.~(\ref{eq:new-formula}) by using the DDP method 
with the deformed contour shown in Fig.~\ref{fig: k plane}, as we sketch in the main text.
The detail of the evaluation is as follows.
We focus on the detoured part (the circular arc around $k_c$), which yields a dominant contribution.
As mentioned
in the main text, we assume that $\bm{d}^{2}\sim i\alpha(k-k_{c})$,
$(\partial_{k}\bm{d})^{2}\sim\beta$, and $(\bm{d}\times\partial_{k}\bm{d})\cdot(\partial_{k}^{2}\bm{d})\sim\eta$
in the vicinity of $k_{c}$. Then $(\partial_{k}\bm{d}\times\bm{d})^{2}=(\partial_{k}\bm{d})^{2}\bm{d}^{2}-(\partial_{k}\bm{d}^{2})^{2}/4\sim\alpha^{2}/4$
leads to 
\begin{align}
\tilde{A}_{+-}\sim\pm\frac{\text{sgn}(\text{Im} k_c)}{4i(k-k_{c})},
\end{align}
where, to avoid confusion, we have introduced $\tilde{A}_{+-}=|A_{+-}|$ [defined as Eq.~(\ref{eq:asym-bc})], which is an analytic function on the complex plane (so that $\tilde{A}_{+-}$ is not necessarily positive real).
This, remarkably, does not depend on any detail ($\alpha,\beta,\eta$, etc.) of the system.  
The sign $\pm$ should be chosen such that $\tilde{A}_{+-}\ge0$ holds on the real axis (typically $+$).
We also expand the exponent as
\begin{align}
\int_{k_{0}}^{k_{1}}dk_{2}\left(\frac{\varepsilon_{+}-\varepsilon_{-}}{eE}+R_{+-}\right)\sim z+\int_{k_{0}}^{k_{c}}dk_{2}\left(\frac{\varepsilon_{+}-\varepsilon_{-}}{eE}+R_{+-}\right),
\end{align}
where
\begin{align}
z=\frac{4}{3}\sqrt{i\alpha}\left(\frac{1}{eE}-\frac{2\eta}{\alpha^{2}}\right)(k_{1}-k_{c})^{3/2}.
\end{align}

The main part of the contour (where the absolute value of the exponential factor is constant) 
extends from $k_c$ to directions where $z$ is real. 
There are three such directions in $2\pi/3$ intervals, as can be seen in Fig.~\ref{fig: k plane}.
The circular arc connects two of them, so it has an angle $2\pi/3$.

We choose either $k_c$ or $k_c^\ast$, such that $\arg z\in[0,\pi]$ holds on the arc \footnote{In typical systems (e.g. the Rice-Mele model), $\alpha>0$ for $\text{Im} k_c>0$ and $\alpha<0$ for $\text{Im} k_c<0$ holds. In such a case we should choose $k_c^{(-\text{sgn}(E))}$. This makes the contour depend on the direction of $E$, as shown in Fig.~\ref{fig: k plane}(a) and (b).}.
Then further we set the infinitesimal radius of the arc
to be $\propto |E|^{2/3-\epsilon}$ with $\epsilon>0$. 
By doing so we can make the contribution of detoured part dominate that of the main part
(for rigorous bounds, see Ref.~\onlinecite{Davis1976}.
The shift vector correction $2\eta/\alpha^{2}$ does not affect the
bounds as it is higher-order w.r.t. $E$).

We then change the integrating variable from $k_{1}$ to $z$. 
The integral (\ref{eq:amp-1st}) along the detoured contour becomes
\begin{align}
a_{+}(t)\sim\pm e^{i\arg A_{+-}(t_{0})}\exp\left[-i\int_{k_{0}}^{k_{c}}dk_{2}\left(\frac{\varepsilon_{+}-\varepsilon_{-}}{eE}+R_{+-}\right)\right]\int_{C}dz\frac{e^{-iz}}{6z},
\label{eq:appendix-zint}
\end{align}
where the contour $C$ is a semicircle with a radius $\propto E^{-\epsilon}\rightarrow\infty$
as $E\rightarrow0$, covering the upper half-plane \footnote{The sign $\pm$ is identical for $k_c$ and $k_c^\ast$ (typically $-$).
This sign is derived from $\tilde{A}_{+-}$ and the direction of the contour.}. 
With the help of Jordan's lemma, the integral along $C$ is given by the residue at $z=0$, i.e., $\oint dze^{-iz}/(6z)=i\pi/3$.
Finally we arrive at Eq.~(\ref{eq:new-formula}) multiplied by $(\pi/3)^2$. 

Actually it is known that the prefactor $\pi/3$
should be corrected by taking into account all order contributions
in the adiabatic perturbation \cite{Davis1976}. To obtain the correct tunneling amplitude,
we need to multiply the $z$ integrand of Eq.~(\ref{eq:appendix-zint}) by $a_{-}(z)$.
$a_{-}(z)$ with the higher-order correction can be obtained as follows:
(i) Differenciating Eq.~(\ref{eq:amp-eom}) with respect to $k$ leads to the differential equation for $a_{-}$,
\begin{equation}
\partial_{k}^{2}a_{-}-\left[\partial_{k}\ln|A_{+-}|+i\left(\frac{\varepsilon_{+}-\varepsilon_{-}}{eE}+R_{+-}\right)\right]\partial_{k}a_{-}+|A_{+-}|^{2}a_{-}=0.
\end{equation}
(ii) In the vicinity of the branching
point $k_{c}$, the differential equation is simplified as 
\begin{align}
36z^2\partial_{z}^{2}a_{-} +36z(1-iz)\partial_{z}a_{-}-a_{-}=0,
\end{align}
which has an asymptotic series solution,
\begin{equation}
a_{-}(z)=1-\sum_{n=1}^{N}\frac{n!i^{n}}{36n^{2}z^{n}}\prod_{k=1}^{n-1}\left(1-\frac{1}{36k^{2}}\right).
\end{equation}

With this solution, the $z$ integral is evaluated as 
\begin{align}
\oint dz\frac{e^{-iz}}{6z}a_{-}(z) & =\frac{i\pi}{3}\left[1-\sum_{n=1}^{N}\frac{1}{36n^{2}}\prod_{k=1}^{n-1}\left(1-\frac{1}{36k^{2}}\right)\right]\\
& =\frac{i\pi}{3}\prod_{k=1}^{N}\left(1-\frac{1}{36k^{2}}\right)\overset{N\rightarrow\infty}{\longrightarrow}2i\sin\frac{\pi}{6}=i
\end{align}
instead of $i\pi/3$. Hence we end up
with Eq.~(\ref{eq:new-formula}) with the prefactor exactly $1$
instead of $(\pi/3)^{2}$.

\begin{figure}[b]
\begin{centering}
\includegraphics[width=\linewidth]{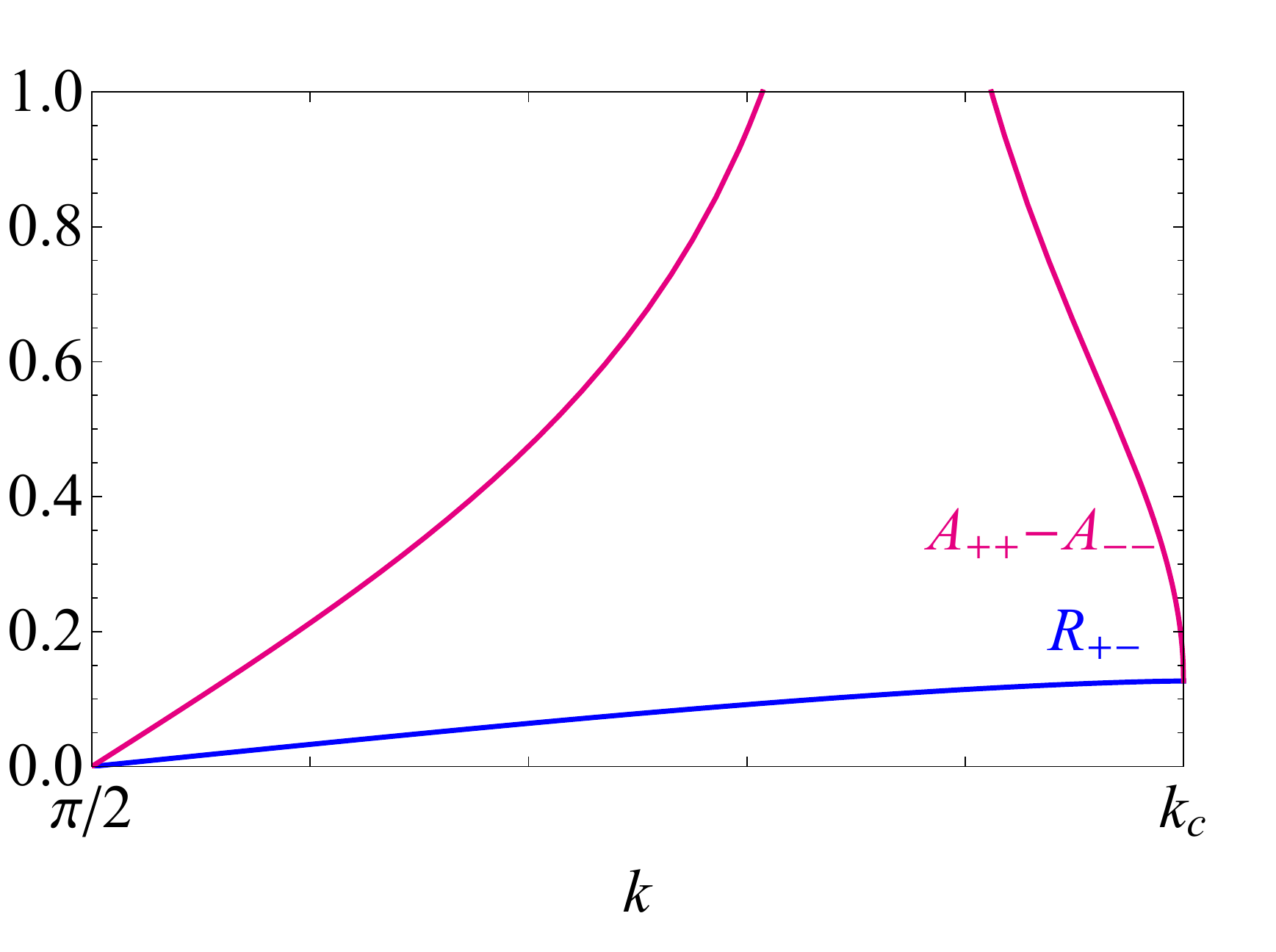}
\par\end{centering}
\caption{\label{fig:appendix} Comparison of the integral $\text{Im}\int_{\pi/2}^{k}dk_2 R_{+-}(k_2)$ with Eq.~(\ref{eq:shift-vector}) (blue) and $\text{Im}\int_{\pi/2}^{k}dk_2 (A_{++}(k_2)-A_{--}(k_2))$ with Eq.~(\ref{eq:joye-integrand}) (magenta) for the Rice-Mele model.}
\end{figure}

\subsection{Comparison of Eq.~(\ref{eq:new-formula}) with the method in Refs.~\onlinecite{Berry1990,Joye1991-1}.\label{sec:comparison}}
The formula for the tunneling amplitude in Refs.~\onlinecite{Berry1990,Joye1991-1} 
can be reproduced by evaluating Eq.~(\ref{eq:new-formula}) with eigenvectors
\begin{align}
|\pm,k(t)\rangle=\frac{(\sqrt{\bm{d}^{2}}\pm\bm{d}\cdot\bm{e}_{z})|\uparrow\rangle\pm\bm{d}\cdot(\bm{e}_{x}+i\bm{e}_{y})|\downarrow\rangle}{\sqrt{2(\bm{d}^{2}\pm\sqrt{\bm{d}^{2}}\bm{d}\cdot\bm{e}_{z})}}.
\end{align}
For this gauge choice, we obtain
\begin{gather}
A_{++}-A_{--}=\frac{((\bm{d}\times\partial_{k}\bm{d})\cdot\bm{e}_{z})(\bm{d}\cdot\bm{e}_{z})}{\sqrt{\bm{d}^{2}}(\bm{d}\times\bm{e}_{z})^{2}},\label{eq:joye-integrand}\\
\arg A_{+-}=\frac{\pi}{2}-\tan^{-1}\frac{2\sqrt{\bm{d}^{2}}(\bm{d}\times\partial_{k}\bm{d})\cdot\bm{e}_{z}}{(\bm{d}\cdot\bm{e}_{z})\partial_{k}\bm{d}^{2}-2(\partial_{k}\bm{d}\cdot\bm{e}_{z})\bm{d}^{2}}.
\end{gather}
If we assume that $\bm{d}(k_{c})\cdot\bm{e}_{z}\neq0$ (in accordance with Ref.~\cite{Joye1991-1}), 
$\arg A_{+-}(k_{c})=\pi/2$ holds in general cases, so that $\text{Im}\int_{k_{0}}^{k_{c}}dk\partial_{k}\arg A_{+-}=0$.
Hence, we can replace $R_{+-}$ in Eq.~(\ref{eq:new-formula}) by $A_{++}-A_{--}$, 
which leads to the expression given in Refs.~\cite{Berry1990,Joye1991-1}.
While this expression is obtained in a particular gauge, 
our formula (\ref{eq:new-formula}) is gauge-invariant and thus free from the assumption $\bm{d}(k_{c})\cdot\bm{e}_{z}\neq0$.
Note that the integrand (\ref{eq:joye-integrand}) has no physical meaning while the integrated value does.
In particular, it diverges as $\sim(k-k_{c})^{-1/2}$ as $k\rightarrow k_c$, so that not suitable for numerical evaluation.
Let us compare the two formulae,
$\text{Im}\int_{\pi/2}^{k}dk_2 R_{+-}(k_2)$ and 
$\text{Im}\int_{\pi/2}^{k}dk_2 (A_{++}(k_2)-A_{--}(k_2))$ with Eq.~(\ref{eq:joye-integrand}),
with a concrete example.
We plot these integrals for the Rice-Mele model Eq.~(\ref{eq:RM}) with $(\delta t,m)=(0.4t,0.4t)$ in Fig.~\ref{fig:appendix}.
While the integrals terminated at a generic $k$ do not coincide, they do at the branching point $k_c$.

\bibliographystyle{apsrev4-1}
\bibliography{reference}

\end{document}